\documentclass{article}

\usepackage{arxiv}

\usepackage[utf8]{inputenc} 
\usepackage[T1]{fontenc}    
\usepackage{hyperref}       
\usepackage{url}            
\usepackage{booktabs}       
\usepackage{amsfonts}       
\usepackage{nicefrac}       
\usepackage{microtype}      
\usepackage{lipsum}
\usepackage{multirow}
\usepackage[font=scriptsize]{subcaption}
\usepackage{cite}
\usepackage{amsmath}
\usepackage{graphicx}
\usepackage{subcaption}
\usepackage{float}

\title{EEG source localization analysis in epileptic children during a visual working-memory task.}

\author{
  Galaris Evangelos\\
  Dipartimento di Matematica e Applicazioni “Renato Caccioppoli”\\
  Universita’ di Napoli Federicco II\\
  Naples, Italy \\
   \\
   \And
 Gallos Ioannis \\
  School of Applied Mathematics and Physical Sciences\\
  National Technical University of Athens\\
  Athens, Greece \\
   \\
   \And
 Myatchin Ivan \\
  Department of Anesthesiology\\
  Sint-Trudo Regional Hospital\\
  Sint-Truiden, Belgium \\
   \\
   \And
Lagae Lieven \\
  Department of Development and Regeneration\\
  Section Paediatric Neurology, KULeuven\\
  Leuven, Belgium \\
   \\
   \And
 Siettos Constantinos \\
  Dipartimento di Matematica e Applicazioni “Renato Caccioppoli”\\
  Universita’ di Napoli Federicco II\\
  Naples, Italy \\
  (Corresponding author, constantinos.siettos@unina.it)
   \\
}

\begin{document}
\maketitle

\begin{abstract}
We localize the sources of brain activity of children with epilepsy based on EEG recordings acquired during a visual discrimination working memory task.  For the numerical solution of the inverse problem, with the aid of age-specific MRI scans processed from a publicly available database, we use and compare three regularization numerical methods, namely the standarized Low Resolution Electromagnetic Tomography (sLORETA), the weighted Minimum Norm Estimation (wMNE) and the dynamic Statistical Parametric Mapping (dSPM). We show that all three methods provide the same spatio-temporal patterns of differences between epileptic and control children. In particular, our analysis reveals statistically significant differences between the two groups in regions of the Parietal Cortex indicating that these may serve as ``biomarkers" for diagnostic purposes and ultimately localized treatment.
\end{abstract}


\section{Introduction}
\paragraph{}Epilepsy affects more than 65 million people worldwide while, approximately 1 out of 150 children is diagnosed with epilepsy during the first 10 years of their life \cite{Aaberg2017}. Although many children self-heal before adulthood, it has been shown that children with epilepsy confront various cognitive and behavioural problems such as problems in learning, attention and memory capacity \cite{Davies2003}. Thus, the systematic study of the brain (dys)functionalities of children with epilepsy, and ultimately the development of efficient/targeted treatments is one of the most challenging problems in neuroscience and beyond. Towards this aim, non-invasive neuroimaging techniques and in particular electroencephalograph (EEG) recordings are commonly used for clinical assessment \cite{Goodin1990, Smith2005, Jan2001, Krause2008, Myatchin2009, Protopapa2014, Yang2012, vanDiesen2013}. However, an analysis at the scalp level does not give insight to the malfunctioning of the actual brain regions and/or their connectivity. On the other hand, fMRI analysis can provide a better insight but it is limited by its low-time resolution. Thus, source localization, i.e. the identification of brain regions from scalp/non-invasive recordings (usually EEG or MEG) has emerged a promising  approach that can facilitate the analysis of brain activity as a clinical diagnostic tool \cite{Plummer2008,Mgevand2018}. However, the source localization problem is an ill-defined problem and as such, it poses open questions regarding its robustness and in general the validity of the obtained results \cite{Kaiboriboon2012}. Thus, comparative studies between the various numerical methods that aspire to solve the source localization problem are critical\cite{Ebersole2010,Moeller2013}. Toward this aim, Jatoi et al.\cite{Jatoi2014} have compared the standardized low resolution brain electromagnetic tomography  (sLORETA) with the exact LORETA (eLORETA) based on EEG recordings of a visual experiment on healthy subjects. Cincotti et al. \cite{Cincotti_2004} compared two techniques for source localization, namely the surface Laplacian and LORETA using EEG recordings from a group of Alzheimer disease patients and age-matched controls. Yao and Devald \cite{Yao_2005} compared the performances of several source localization methods on the basis of both simulated and experimental EEG data of somatosensory evoked potentials. Attal and Schwartz \cite{Attal_2013} compared the performance of three methods, namely the weighted minimum norm (wMNE), sLORETA and the dynamic statistical parameter mapping (dSPM) for the  characterization of distortions in cortical and subcortical regions using a realistic anatomical and electrophysiological model of deep brain activity. Seeland et al. \cite{Seeland_2018} compared wMNE, sLORETA and dSPM using EEG data taken from eight subjects performing voluntary arm movements.

Regarding epilepsy, the majority of the studies have performed source localization with the aid of EEG-fMRI recordings and/or simulated data approximating epileptic spatio-temporal patterns such as spikes and discharges. For example, Ioannides et al. \cite{Ioannides_2015} assessed the performance of two source localization methods, wMNE and eLORETA using MEG signals of ictal and interictal epileptiform discharges in epilepsy and K-complexes. Chowdhury et al.\cite{Chowdhury_2016} compared the performance of the coherent Maximum Entropy on the Mean (cMEM) and the 4th order Extended Source Multiple Signal Classification (4-ExSo-MUSIC) using MEG and EEG synthetic signals mimicking normal background and epileptic discharges. Hasan et al. \cite{Hassan_2016} evaluated four algorithms (dSPM, wMNE, sLORETA and cMEM) using simulated data from a combined biophysical/physiological model used to generate interictal epileptic spikes as well as real EEG data recorded from one epileptic patient who underwent a full presurgical evaluation for drug-resistant focal epilepsy. Moeller et al. \cite{Moeller2013} provides a review of the studies that used EEG-fMRI recordings to assess different types of epileptic form activity, underpinning the necessity for comparing with other methods including EEG source analysis.

Fewer studies have dealt with source-level analysis and compared different source localization methods using EEG clinical data taken by children with epilepsy. Among these studies, Adebimpe et al. \cite{Adebimpe2016} performed source localization using eLORETA to investigate changes in functional connectivity in children with Benign rolandic epilepsy with centrotemporal spikes using resting-state EEG recordings. Groening et al. \cite{Groening2009} combined EEG–fMRI and EEG source analysis to identify epileptogenic foci in children. Elshoff et al. \cite{Elshoff2012} examined the efficiency of EEG‐fMRI and EEG source analysis to localize the point of seizure onset in children with refractory focal epilepsy.\\ 
The above studies have focused mainly on the study of brain regions that are activated  during seizure periods or before their onset. Several other studies have also aimed at analyzing the emerged patterns during seizure periods. For example, Fergus et al. \cite{Fergus2015} used a supervised machine learning approach to classify seizure and non-seizure records using an open dataset of seizured EEG signals from both children and adults.

On the other hand, it has been shown, that studying epileptic seizure-free EEG recordings is of great importance as such analysis can facilitate the identification of patients at risk of epilepsy and/or forecast forth-coming seizures (for a discussion and review of ictal and interictal activity and their analysis see for example \cite{vanDiesen2013}.

Here, we perform a source-localization analysis of the brain activity of well-controlled epileptic children during a visual Working Memory (WM) task. WM is commonly viewed as a functional integration system with limited capacity that is able to store information within a short-term register and simultaneously manipulate it on-line. Thus, WM is one of the most important components of information processing and its dysfunction leads to various problems in several cognitive functions including mental arithmetic \cite{Swanson2006}, reading \cite{Gathercole2004, Gathercole2008}, decision making \cite{Toth2003} and reasoning \cite{Ruff2003}. Epilepsy affects a lot the WM functioning as it has been shown by many studies \cite{Gulgonen2000, Chaix2006, Campo2009, Wagner2009, Luton2010}. 

Here, for the solution of the inverse problem, we use three methods, namely, sLORETA \cite{Pasqualmarqui2002}, dSPM \cite{Dale2005} and wMNE \cite{Iwaki1998}. A statistical comparative analysis between methods and groups (healthy children vs children with epilepsy) revealed the crucial role of the Superior Parietal Lobule (SPL) and Inferior Parietal Lobule (IPL) at WM. Our findings are in line with  fMRI studies \cite{Otsuka2008, Koenigs2009, Olson2009, Ravizza2005} that have shown that SPL and IPL are being involved in WM processing and thus can serve as a ``biomarker" for identifying, monitoring and assessing epilepsy in children.

\section{Materials and methods}
\subsection{Experimental Procedure}

\subsubsection{Subjects}
\paragraph{}In the study group, 21 children with established childhood epilepsy (age 6—16 years old; mean 11.43 years, SD $\pm$2.3, 10 boys) were enrolled. These children were diagnosed with one out of two following epilepsy syndromes: benign rolandic epilepsy (BRE) (n = 9) and idiopathic generalized epilepsy (GE) (n = 12, including childhood absence epilepsy (n = 5) and generalized epilepsy with tonic-clonic seizures (n = 7)). All children were admitted to the neurophysiology laboratory of the University Hospital of Leuven for a 24-h video-EEG monitoring during which the ERP study was done. They had no anti-epileptic treatment (n = 3) or were on standard anti-epileptic medication (monotherapy, n = 15, duotherapy, n = 3), with drug dosages always being within normal ranges. Patients on monotherapy received valproic acid (n = 7), carbamazepine (n = 4), lamotrigine (n=3) or sulthiame (n = 1). Patients on duotherapy received different drug combinations \cite{Myatchin2009}. None of the patients had structural brain abnormalities; in 18 patients, brain MRI was performed showing normal findings in all cases. Only patients with at least an eight days seizure-free period preceding the test were included \cite{Myatchin2009}. Thereby we could avoid an acute effect of epileptic seizures on the child’s performance. All children followed mainstream school and none had a history of learning problems. 
As a control group, 25 age-matched non-epileptic children (mean age 10.76 years, SD $\pm$3.4, 17 boys) were selected, who did not have any school problem either.
The study protocol was approved by the Ethical Committee of University Hospital of Leuven. For more details about the experimental procedure please see Myatchin et al. \cite{Myatchin2009}.

\subsubsection{Design and Stimuli}
\paragraph{}The event-related potentials study was done as part of video-EEG monitoring. A visual one-backmatching working memory task was performed: children observed a continuous stream of seven different figures presented one after the other in pseudorandom order at the middle of a computer monitor, which was located at a distance of 1.0m from the subject’s eyes. Everyday figures were used (horse, wardrobe, jacket, cake, comb, bunch of grapes, hammer), white with a black contour on grey background, size 7.5 cm $\times$ 6.5 cm, visual angle $4^\circ18'\times3^\circ02'$. Each stimulus was presented for 1.5 s, followed by a delay of 1.0 s, after which the next stimulus was presented. During the delay period a fixation point (dark-grey cross) was shown at the middle of the screen to facilitate eyes fixation. Any figure identical to the one immediately preceding it was defined as a target stimulus (probability 0.30). Children were asked to respond to all targets by pressing a button with their dominant hand. Both accuracy and speed were stressed. The single experimental block contained 120 trials, 36 of which were targets. The duration of the block was 5 min. This is an easy working memory task, which was chosen to ensure a good level of participant’s performance. 

First, the electrode placement and impedance calibration was performed. After that, the experimental procedure was described to the child. The child was seated comfortably in a dimly lit registration room and was instructed to look at the middle of the computer screen placed in front of him to avoid unnecessary eye movements; a fixation point (dark-grey cross) was shown between figures to facilitate eye fixation. The child was also instructed to avoid movements to reduce muscle artifacts in the EEG signal. The instruction for the task was given directly before the task. During the experiment, no interaction with the experimenter was allowed during the task and the experimenter sat out of sight of the child.\\

\subsubsection{EEG recordings}
\paragraph{}Nineteen Ag/AgCl electrodes (Technomed Europe) were placed according to the international 10-20 system at Fp1, Fp2, F3, F4, F7, F8, Fz, C3, C4, Cz, T3, T4, T5, T6, P3, P4, Pz, O1 and O2. Placement of additional four EOG electrodes resulted in two EOG channels: horizontal EOG – two electrodes on the outer canthi of eyes, and vertical EOG – two electrodes above and below one eye. EOG channels allowed us to detect both vertical and horizontal eye movements in order to effectively remove them from EEG recording during subsequent preprocessing of the signal (see below). Two linked mastoid electrodes were used as a reference. EEG was sampled at a frequency of 1000 Hz with 12 bits A/D converter and amplified using a band-pass filter of 0.095 – 70 Hz. Notch filter was off. Registration of the digital EEG was made using the software program BrainlaB 4.0 (OSG, Belgium). The impedance of all electrodes was monitored for each subject prior to recording and was always kept below 5 k$\Omega$.\\

\subsubsection{Data pre-processing}
\paragraph{}Data pre-processing was performed offline using the EEGLAB v.5.02 toolbox (Matlab 7.0.4 platform) \cite{Delorme2004}. The ECG channel was factored out. Data were filtered with a 50 Hz digital low pass filter. Eye movement artifacts were marked and removed from the continuous signal without affecting the signal itself using an ICA-based algorithm \cite{Myatchin2009}. EEG fragments containing movement artifacts as well as any epileptic activity were removed based on visual inspection of the data. This resulted in an EEG signal clean from (eye) movement artifacts and epileptic activity, which was then used for further analysis. 
Afterwards, the continuous EEG signal was epoched according to the type of stimulus (Target and Non Target), with 200 ms pre-stimulus (delay period) and 400 ms poststimulus (presentation period of the second stimulus, where the motor responses had not yet taken place). Omitted Target trials (i.e. trials without correct motor response) and committed Non Target trials (i.e. trials with a wrong motor response) were excluded from the analysis. We then performed a down-sampling at 500 Hz and we applied a baseline correction by subtracting the mean value of the 200 ms of the pre-stimulus period. Overall, we ended up with 92 datasets (21 epileptic $\times$ 2 trial types + 25 control $\times$ 2 trial types) of 19 multi timeseries, which were divided into four group types (Epileptic- Target (ET), Epileptic-Non Target (ENT), Control-Target (CT), Control-Non Target (CNT)).

\subsection{Numerical Solution of the Source localization Problem}

\paragraph{} Source localization aims at identifying  the (unknown) sources of the brain from data taken usually from noninvasive electromagnetic recording (here: EEG recordings). Its solution involves a forward and an inverse problem. The forward problem refers to the calculation of the electric potentials of the electrodes starting from a given electrical source. The solution of the forward problem is related to the construction of a head model. The head model contains both anatomical information and the conductivities of three layers, namely  the skull, the cortex and the scalp \cite{Kybartaite2012}. Anatomical images can be obtained experimentally with the aid of Magnetic Resonance Imaging (MRI) scans, while volume conduction models can be constructed using e.g. the Boundary Element Method (BEM) \cite{Fuchs2001} or the Finite Elements Model method (FEM)\cite{Wolters2007}. The head model volume is tessellated into small-sized cubes, the voxels. Sources may be associated to single voxels or clusters of voxels. Here, each voxel is asscociated to a single source. The relation between the scalp recordings and the discretized head model volume is performed using the linear matrix equation:\\

\begin{equation}
  \boldsymbol{V}=\boldsymbol{G}\boldsymbol{x}+\boldsymbol{\epsilon},
  \label{eq1}
\end{equation}

where $\boldsymbol{V}$  is a known $N \times 1$ matrix which contains the time instances as recorded by each channel ($N$ is the number of channels), $\boldsymbol{x}$ is the unknown $M \times 1$ matrix of the intensities of the M sources ($M$ is the number of voxels).\\
The matrix $\boldsymbol{G}$, with dimensions $N \times M$ is the so-called lead field matrix that contains the information of the head geometry and conductivities. $\boldsymbol{G}$ is known (from the solution of the so called forward problem (see e.g. in \cite{Hallez2007})) and is related with the head model \cite{Marqui1993}; $\epsilon$ reflects the noise in the measurements.\\
The inverse problem is ill-defined, as there is an infinite number of combinations of positions and intensities that could effectively produce the electric potentials and magnetic fields measured. The general idea behind its solution is to express it as a linear optimization problem with regularization:\\

\begin{equation}
 \hat{x}=\min_{x}(||V-Gx||^2_2+\sum^k_{i=1}a_i||W_ix||_p)
 \label{eq2}.
\end{equation}

In the above, $k$ is the number of regularization constraints (refelcting the a-priori physiological information); the matrix $W$, $M \times M$, is a weighted matrix related to the imposed constraints; $\alpha_i$ is the regularization parameter and denote the importance of every constraint.\\
For different choices of $W$, $k$ and $p$ (reflecting the type of the norm), we get different methods.\\
Here, for our analysis, we used and compared three different \cite{Seeland2018} methods, namely the weighted Minimum Norm Estimation (wMNE), the standarized Low Resolution Electromagnetic Tomography (sLORETA) and the dynamic Statistical Parametric Mapping (dSPM)  that are described below.

\subsubsection{Weighted Minimum Norm Estimation}
\paragraph{} For $W=I$ (the identity matrix) and $p=2$ (the $L-2$ norm) in \ref{eq2} we get the Minimum Norm Estimation (MNE) \cite{Hamalainen1994}. MNE uses the mathematical assumption that the best solution, through the infinite set of solutions, is the one with the minimum norm. Despite the fact that MNE was the first method used to extract a 3D distributed solution, the simplicity of its assumption often leads to inadequate solutions. In particular, it has been shown, that this method fails in identifying deep sources \cite{Stenroos2013}. Because of the minimum norm constraint, sources that are located in deep regions are moved closer to the cortex.\\
The wMNE method is a variation of the MNE that improves the problem of the mislocation of the deep sources. wMNE uses instead of the identity matrix, a diagonal matrix $W_c$ that contains the weighting factors. From the multiple choices that can be chosen as weighted factors, usually $W_c=diag(||G_i||_2)$ (for i =1,...,M) is chosen \cite{Grech2008}. Then, the unique solution is  given by:\\
\begin{equation}
    x_{wMNE}=LV,
    \label{eq3}
\end{equation}
where $L=G^{T}(G^TG+\alpha W_c)^{-1}$ is called the inverse operator with dimensions ({$M\times N$}).

\subsubsection{The Dynamic Statistical Parametric Mapping}
\paragraph{} The Dynamic Statistical Parametric Mapping (dSPM) \cite{Dale2005} is similar to the wMNE but uses a different regularization.  dSPM computes the source estimates of the noise based on the noise covariance matrix $C_{\epsilon}=\alpha H$ and normalizes the rows of the inverse operator.\\ $H=I-\frac{1^{T}1}{1^{T}1}$ is the centering matrix and plays the role of the identity matrix in the measurement space. Then, from equation \ref{eq3}, the source estimates of the noise form a diagonal matrix:\\

\begin{equation}
    C_{\hat{x}}=W_{dSPM}=LC_{\epsilon}L^T.
    \label{eq4}
\end{equation}
Thus, the dSPM solution is given by:
\begin{equation}
    x_{dSPM}=L_{dSPM}V,
    \label{eq5}
\end{equation}
where $L_{dSPM}=W_{dSPM}L$.

\subsubsection{Standarized Low Resolution Electromagnetic Tomography}
\paragraph{} sLORETA considers another source of variance, except from the covariance of the measurement noise $C_{\epsilon}$: the covariance of the actual sources $C_{x}=I$. Assuming that the activity of the actual sources and the noise of the measurements are uncorrelated and based on the linear relation of equation \ref{eq1}, we have:\\
\begin{equation}
    C_{V}=GC_{x}G^{T}+C_{\epsilon}=GG^{T}+\alpha H.
    \label{eq6}
\end{equation}

Substituting equation \ref{eq6} to \ref{eq3}, and taking into account the linear relation of equation \ref{eq3}, we can estimate the variation of the estimated sources as:\\

\begin{equation}
    C_{\hat{x}}=LC_{V}L^{T}=L(GG^{T}+\alpha H)L^{T}=G^{T}(GG^{T}+\alpha H)^{-1}G.
    \label{eq7}
\end{equation}

The covariance of the estimated sources is equivalent to the Backus and Gilbert resolution matrix \cite{Backus1968}, which is given by plugging equation \ref{eq1} into \ref{eq3} and substituting the inverse operator to get:\\

\begin{equation}
    \hat{x}=LGx=G^{T}(GG^{T}+\alpha H)Gx=Ax=C_{\hat{x}}x,
    \label{eq8}
\end{equation}
where $A=LG$ is the resolution matrix.

In this case, the solution is given by:
\begin{equation}
    x_{sLORETA}=L_{sLORETA}V,
\end{equation}
where $L_{sLORETA}=AL$.

\subsubsection{Head models for children}
In our study, we did not have individual MRI scans for each child that participated to the experiment. Thus, in the absence of such specific information, we used age-specific MRI templates for children acquired from the ``Neurodevelopmental MRI database" \cite{Richards2016, Evans2006, Sanchez2012, Richards2015}. The goal of this database is to provide for research purposes, exactly in the absence of specific MRI scans, a series of age-appropriate average MRI reference templates and related information. Each template was constructed using identical procedures to facilitate comparisons across lifespan. The database consists of average templates (T1W and T2W), segmenting priors, and stereotaxic atlases \cite{Richards2016}. The ``Neurodevelopmental MRI Database” is available online (\url{http://jerlab.psych.sc.edu/NeurodevelopmentalMRIDatabase/}). The data-base is publicly available to researchers upon request for clinical and experimental studies of normal and pathological brain development. The data is shared under a Creative Commons Attribution-NonCommercial-Noderivs 3.0 Unported License (CC BY-NC-ND 3.0; \url{http://creativecommons.org/licenses/by-nc-nd/3.0/deed.en$_$US}).\\
Using this database, we were able to construct an ``average" age-specific head model for each child taking into account its age. For our study, we constructed 11 averaged head models (taking into account the database with head models of children between 6 and 16 years old, i.e. one ``average" head model per year). In table \ref{table1}, we provide information about the total number of MRI scans per age.

\begin{table}[h]
\centering
\begin{tabular}{c rrrrrrrrrrr}
\hline
\textbf{Age} &6 &7 &8 &9 &10 &11 &12 &13 &14 &15 &16\\
\hline
\textbf{1.5T} &27 &27 &46 &46 &62 &31 &37 &34 &32 &32 &34\\
\textbf{3.0T} &10 &  &19 & &16 & &15 &11 &30 & &13\\
\hline
\textbf{Combined} &37 &27 &56 &46 &72 &31 &47 &34 &42 &32 &44\\
\hline
\end{tabular}
\caption{ Total number of scans per age for 1.5T, 3.0T and combined average MRI templates. All 1.5T MRIs and part of 3.0T MRIs are included in the "Combined" column as in the original publications \cite{Richards2016, Sanchez2012}}
\label{table1}
\end{table}

Here, for the construction of the head models, as skull conductivities are age-dependent \cite{Wendel2010}, we used different conductivities ratios (CR, cortex/skull) for every age-dependent model. The conductivity value for scalp and cortex was set to the standard value of 0.33 S/m \cite{McCann2019}. Table \ref{table2} presents analytically the different conductivity ratios for every age \cite{McCann2019}.

\begin{table}[h]
\centering
\begin{tabular}{c rrrrrrrrrrr}
\hline
\textbf{Age} &6 &7-8  &9-10  &11-12  &13-14 &15-16 \\

\textbf{CR} &15 &20  &30  &40 &50  &60 \\
\hline
\end{tabular}
\caption{ Conductivity ratios (cortex/skull) for every age-dependent head model. The standard conductivity value for scalp and cortex was set to 0.33 S/m \cite{McCann2019}.}
\label{table2}
\end{table}

\section{Results}
\paragraph{}For our analysis, we used the BrainStorm toolbox for matlab \cite{Baillet1999}. The source-reconstructed time series were obtained by combining the EEG recordings with the appropriate (respect to the age of the subject) constructed MRI templates. From each template, we extracted three layers (scalp, inner skull, outer skull) and the source space (cortical surface). The number of vertices for each layer were set to 2562 vertices for each surface. Then, the volume conduction models were constructed in openMEEG software \cite{Gramfort2010} which uses the BEM. The space resolution for the source model was set to 5124 voxels with fixed orientation perpendicular to the cortex surface.\\ 
Thus, the time series at the source level were reconstructed using  wMNE, dSPM and sLORETA. The noise was computed from the raw EEG data using the pre-stimulus period for baseline correction and then the noise covariance matrix was calculated. A  parameter that has to be determined is the ``signal to noise ratio" (SNR). In Brainstorm, the computation of SNR is performed as in the original MNE software of Hamalainen \cite{Gramfort2013}. The signal covariance matrix is ``whitened" by the noise covariance matrix and the square root of the mean of its spectrum  yields the average amplitude of SNR. The default value in Brainstorm is set to 3.

The main results of source localization procedure are presented analytically at table \ref{source results}. For our illustrations, we have split the time period to three main intervals: the pre-stimulus period [-200ms 0ms), the period exactly after the stimulus [0ms - 199ms] and the post-stimulus period [200ms - 400ms). Our analysis reveals similar results when applying the different methods.\\
In Table \ref{numerics} we also provide the numerical residuals for each method (the L2-norm ($res=||V-G\hat{x}||_2$, where $G$ is the forward operator and $\hat{x}$ the estimated amplitudes), as well as the corresponding values of the regularization terms.

\begin{table}[h!]
\centering
\resizebox{\columnwidth}{!}{%
\begin{tabular}{llllll}
\textbf{Time period} & \textbf{Method} & \textbf{CT} & \textbf{CNT} & \textbf{ET} & \textbf{ENT} \\ \hline
\multicolumn{1}{l|}{\multirow{3}{*}{\textbf{\begin{tabular}[c]{@{}l@{}}pro-stimulus\\ (-200ms - -1ms)\end{tabular}}}} & \multicolumn{1}{l|}{wMNE} & \begin{tabular}[c]{@{}l@{}}-Right occipital\\ lobe ($\sim$140 voxels)\end{tabular} & \begin{tabular}[c]{@{}l@{}}-Occipital lobe\\ ($\sim$100 voxels)\end{tabular} & \begin{tabular}[c]{@{}l@{}}-Right occipital\\ lobe ($\sim$90 voxels)\end{tabular} & \begin{tabular}[c]{@{}l@{}}-Occipital lobe\\ ($\sim$80 voxels)\end{tabular} \\ \cline{3-6} 
\multicolumn{1}{l|}{} & \multicolumn{1}{l|}{dSPM} & \begin{tabular}[c]{@{}l@{}}-Right occipital\\ lobe ($\sim$220 voxels)\end{tabular} & \begin{tabular}[c]{@{}l@{}}-Right occipital\\ lobe ($\sim$150 voxels)\end{tabular} & \begin{tabular}[c]{@{}l@{}}-Right occipital\\ lobe ($\sim$160 voxels)\end{tabular} & \begin{tabular}[c]{@{}l@{}}-Occipital lobe\\ ($\sim$190 voxels)\end{tabular} \\ \cline{3-6} 
\multicolumn{1}{l|}{} & \multicolumn{1}{l|}{sLORETA} & \begin{tabular}[c]{@{}l@{}}-Right occipital\\ lobe\\ ($\sim$530 voxels)\end{tabular} & \begin{tabular}[c]{@{}l@{}}-Occipital lobe\\ ($\sim$230 voxels)\\ -Left parietal\\ lobe ($\sim$150 voxels)\end{tabular} & \begin{tabular}[c]{@{}l@{}}-Right occipital\\ lobe ($\sim$290 voxels)\end{tabular} & \begin{tabular}[c]{@{}l@{}}-Right occipital\\ lobe ($\sim$270 voxels)\end{tabular} \\ \hline
\multicolumn{1}{l|}{\multirow{3}{*}{\textbf{\begin{tabular}[c]{@{}l@{}}exactly after stimulus\\ 0ms - 199ms\end{tabular}}}} & \multicolumn{1}{l|}{wMNE} & \begin{tabular}[c]{@{}l@{}}-Occipital lobe\\ ($\sim$120 voxels)\end{tabular} & \begin{tabular}[c]{@{}l@{}}-Right occipital\\ lobe ($\sim$100 voxels)\end{tabular} & \begin{tabular}[c]{@{}l@{}}-Occipital lobe\\ ($\sim$90 voxels)\end{tabular} & \begin{tabular}[c]{@{}l@{}}-Superior parietal\\ lobe ($\sim$50 voxels)\\ -Occipital lobe\\ ($\sim$110 voxels)\end{tabular} \\ \cline{3-6} 
\multicolumn{1}{l|}{} & \multicolumn{1}{l|}{dSPM} & \begin{tabular}[c]{@{}l@{}}-Right occipital\\ lobe ($\sim$330 voxels)\end{tabular} & \begin{tabular}[c]{@{}l@{}}-Right occipital\\ lobe ($\sim$310 voxels)\end{tabular} & \begin{tabular}[c]{@{}l@{}}-Occipital lobe\\ ($\sim$240 voxels)\end{tabular} & \begin{tabular}[c]{@{}l@{}}-Superior parietal\\ lobe ($\sim$100 voxels)\\ -Right occipital\\ lobe ($\sim$330 voxels)\end{tabular} \\ \cline{3-6} 
\multicolumn{1}{l|}{} & \multicolumn{1}{l|}{sLORETA} & \begin{tabular}[c]{@{}l@{}}-Right occipital\\ lobe ($\sim$590 voxels)\end{tabular} & \begin{tabular}[c]{@{}l@{}}-Occipital lobe\\ ($\sim$510 voxels)\end{tabular} & \begin{tabular}[c]{@{}l@{}}-Occipital lobe\\ ($\sim$450 voxels)\end{tabular} & \begin{tabular}[c]{@{}l@{}}-Superior parietal\\ lobe ($\sim$240 voxels)\\ -Occipital lobe\\ ($\sim$500 voxels)\end{tabular} \\ \hline
\multicolumn{1}{l|}{\multirow{3}{*}{\textbf{\begin{tabular}[c]{@{}l@{}}post-stimulus\\ \\ 200ms - 399ms\end{tabular}}}} & \multicolumn{1}{l|}{wMNE} & \begin{tabular}[c]{@{}l@{}}-Parietal lobe\\ ($\sim$70 voxels)\end{tabular} & \begin{tabular}[c]{@{}l@{}}-Right parietal\\ lobe ($\sim$60 voxels)\end{tabular} & \begin{tabular}[c]{@{}l@{}}-Right parietal\\ lobe ($\sim$60 voxels)\end{tabular} & \begin{tabular}[c]{@{}l@{}}-Right parietal\\ lobe ($\sim$80 voxels)\end{tabular} \\ \cline{3-6} 
\multicolumn{1}{l|}{} & \multicolumn{1}{l|}{dSPM} & \begin{tabular}[c]{@{}l@{}}-Parietal lobe\\ ($\sim$180 voxels)\end{tabular} & \begin{tabular}[c]{@{}l@{}}-Parietal lobe\\ ($\sim$110 voxels)\end{tabular} & \begin{tabular}[c]{@{}l@{}}-Right parietal\\ lobe ($\sim$200 voxels)\end{tabular} & \begin{tabular}[c]{@{}l@{}}-Parietal lobe\\ ($\sim$190 voxels)\end{tabular} \\ \cline{3-6} 
\multicolumn{1}{l|}{} & \multicolumn{1}{l|}{sLORETA} & \begin{tabular}[c]{@{}l@{}}-Parietal lobe\\ ($\sim$390 voxels)\end{tabular} & \begin{tabular}[c]{@{}l@{}}-Parietal lobe\\ ($\sim$360 voxels)\end{tabular} & \begin{tabular}[c]{@{}l@{}}-Right parietal\\ lobe ($\sim$470 voxels)\end{tabular} & \begin{tabular}[c]{@{}l@{}}-Parietal lobe\\ ($\sim$500 voxels)\end{tabular} \\ \hline
\end{tabular}}%
\caption{Group averaged sources as obtained by the three methods: wMNE, dSPM and sLORETA. CT: Control Target, CNT: Control non-Target, ET: Epileptic Target, ENT: Epileptic non-Target.}
\label{source results}
\end{table}

\begin{table}[h!]
\begin{tabular}{llll}
\textbf{Method} & \textbf{Group} & \textbf{Residuals ($\mu$V)} & \textbf{\begin{tabular}[c]{@{}l@{}}Regularization \\ term (nA-m)\end{tabular}} \\ \hline
\multicolumn{1}{l|}{\multirow{4}{*}{\textbf{wMNE}}} & \multicolumn{1}{l|}{CT} & 4.53$\pm$0.94 & 13.2$\pm$2.14 \\ \cline{3-4} 
\multicolumn{1}{l|}{} & \multicolumn{1}{l|}{ET} & 5.34$\pm$1.02 & 15.41$\pm$3.23 \\ \cline{3-4} 
\multicolumn{1}{l|}{} & \multicolumn{1}{l|}{CNT} & 3.63$\pm$1.25 & 13.13$\pm$3.42 \\ \cline{3-4} 
\multicolumn{1}{l|}{} & \multicolumn{1}{l|}{ENT} & 6.42$\pm$1.76 & 16.31$\pm$3.67 \\ \hline
\multicolumn{1}{l|}{\multirow{4}{*}{\textbf{dSPM}}} & \multicolumn{1}{l|}{CT} & 3.92$\pm$0.71 & 12.78$\pm$3.12 \\ \cline{3-4} 
\multicolumn{1}{l|}{} & \multicolumn{1}{l|}{ET} & 5.21$\pm$1.18 & 15.91$\pm$3.54 \\ \cline{3-4} 
\multicolumn{1}{l|}{} & \multicolumn{1}{l|}{CNT} & 4.11$\pm$1.37 & 14.34$\pm$4.15 \\ \cline{3-4} 
\multicolumn{1}{l|}{} & \multicolumn{1}{l|}{ENT} & 7.08$\pm$1.58 & 17.37$\pm$4.52 \\ \hline
\multicolumn{1}{l|}{\multirow{4}{*}{\textbf{sLORETA}}} & \multicolumn{1}{l|}{CT} & 3.12$\pm$0.59 & 10.91$\pm$3.19 \\ \cline{3-4} 
\multicolumn{1}{l|}{} & \multicolumn{1}{l|}{ET} & 4.04$\pm$0.86 & 11.46$\pm$2.84 \\ \cline{3-4} 
\multicolumn{1}{l|}{} & \multicolumn{1}{l|}{CNT} & 2.98$\pm$0.43 & 11.51$\pm$2.9 \\ \cline{3-4} 
\multicolumn{1}{l|}{} & \multicolumn{1}{l|}{ENT} & 3.45$\pm$0.73 & 12.43$\pm$3.57
\end{tabular}
\caption{Mean and standard deviation values for the L2-norm residuals and the regularization term for each method and each group.}
\label{numerics}
\end{table}

\subsection{Hypothesis Testing Using T-test}

\paragraph{}After the data pre-processing and the implementation of the source-localization algorithms, as described in the previous section, we got 92 time-series at the source space. Our data have a spatio-temporal structure: number of voxels (spatial dimension) and time points (time dimension). In order to perform two-sample T-tests for the identification of statistically significant differences in the activity of the sources among groups, we checked three basic assumptions. In particular, we checked if (1) the amplitude of the source signal at each voxel and at each time instant follows a normal distribution among subjects in a group, (2) the variances of the amplitude of the source signal at each voxel and at each time instant of CT, ET and CNT,ENT are equal, and (3) the amplitudes of the source signal at each voxel and at each time instant are independent for CT, ET and CNT, ENT.

For the test of normality, we used the Shapiro-Wilk test \cite{Ahmed2019}. The null hypothesis of the test is that a sample comes from a normal distribution. The test statistic reads:
\begin{equation}
    W=\frac{(\sum^n_{i=1}\alpha_{i}x_{(i)}{})^2}{\sum^n_{i=1}(x_{i}-\bar{x})^2},
    \label{eq10}
\end{equation}
where $x_{(i)}$ is the i-th order statistic, $\bar{x}$ is the sample mean and coefficients $\alpha_{i}$ are given by $(\alpha_{1},\alpha_{2},...,\alpha_{n})=\frac{m^TV^{-1}}{C}$. C is a vector norm $C=||V^{-1}m||=(m^TV^{-1}V^{-1}m)^{\frac{1}{2}}$, $m=(m_1,m_2,...,m_n)^T$ are the expected values of the order statistics of independent and identically distributed random variables sampled from the standard normal distribution and V is the covariance matrix of those normal order statistics. F-tests were performed to validate the second assumption (i.e. the equality of the variances of the amplitude values of each voxel between ET and CT and between ENT and CNT).

Thus, we tested for normality and equality of variances for each voxel and each time sample (i.e. we have performed a total of $5.148\times600=3.088.800$ t-tests). The level of significance was set to $p<0.05$, meaning  that the risk of taking a false positive is 5\% of the cases.  Here, in order to deal with the multicomparison problem, we used the false discovery rate (FDR) correction \cite{Benjamini1995}. Following this procedure, the null hypothesis could not be rejected. Similarly, the F tests validated also the second assumption. The independence is reasonably assumed to hold true. 

\begin{figure}[h!]
  \centering
    \begin{subfigure}[b]{0.46\textwidth}
    \includegraphics[width=\linewidth]{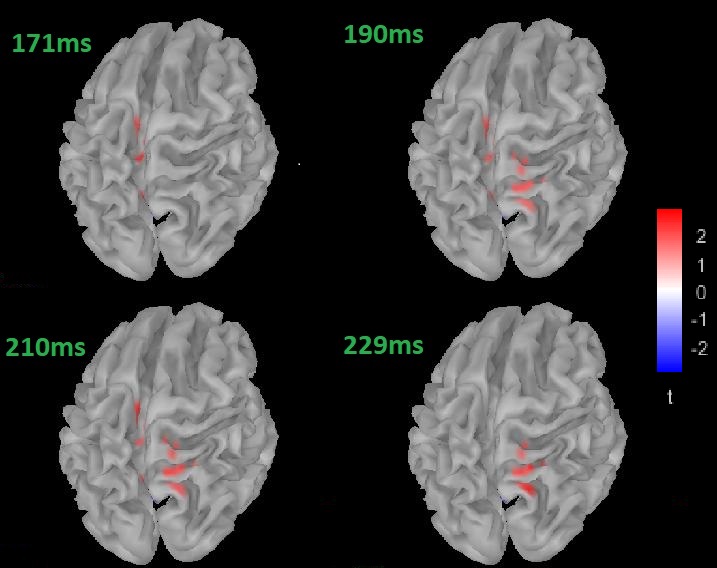}
     \caption{\normalsize{wMNE}}
  \end{subfigure}
  \begin{subfigure}[b]{0.45\textwidth}
    \includegraphics[width=\linewidth]{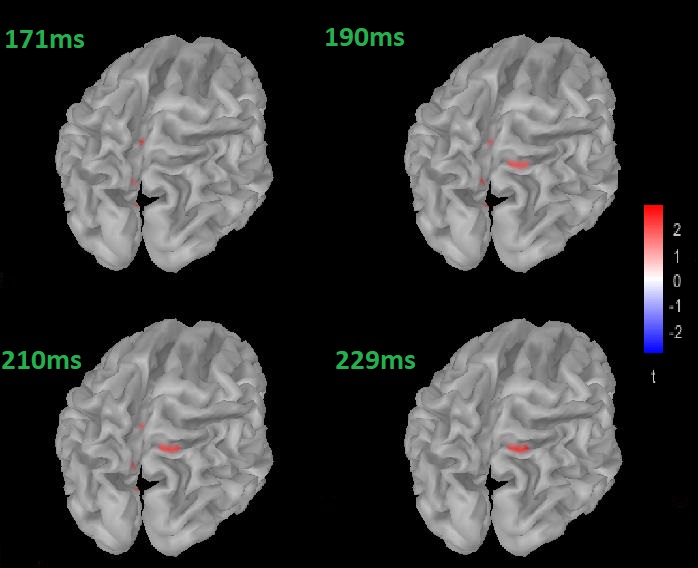}
    \caption{\normalsize{dSPM}}
  \end{subfigure}
  \begin{subfigure}[b]{0.45\textwidth}
    \includegraphics[width=\linewidth]{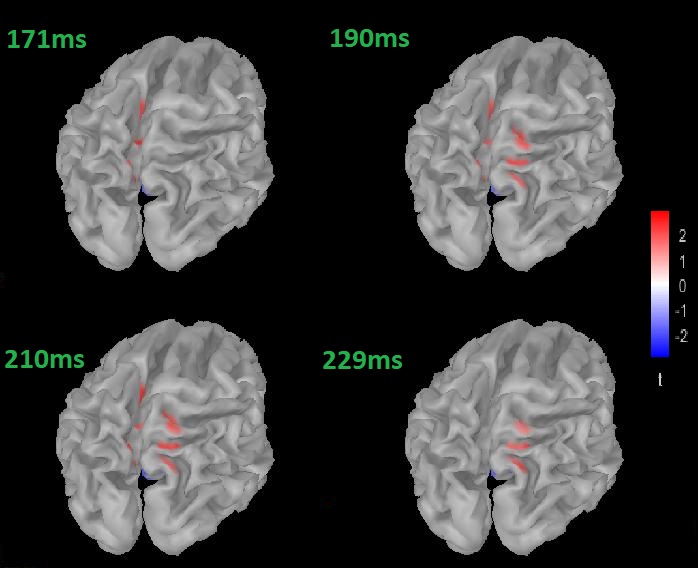}
    \caption{\normalsize{sLORETA}}
  \end{subfigure}
  \caption{ENT vs CNT: SPL mainly of the right hemisphere activate more for the ENT group at the time interval from 170 to 230 ms.}
  \label{figure1}
\end{figure}

\begin{figure}[h!]
  \centering
    \begin{subfigure}[b]{0.45\textwidth}
    \includegraphics[width=\linewidth]{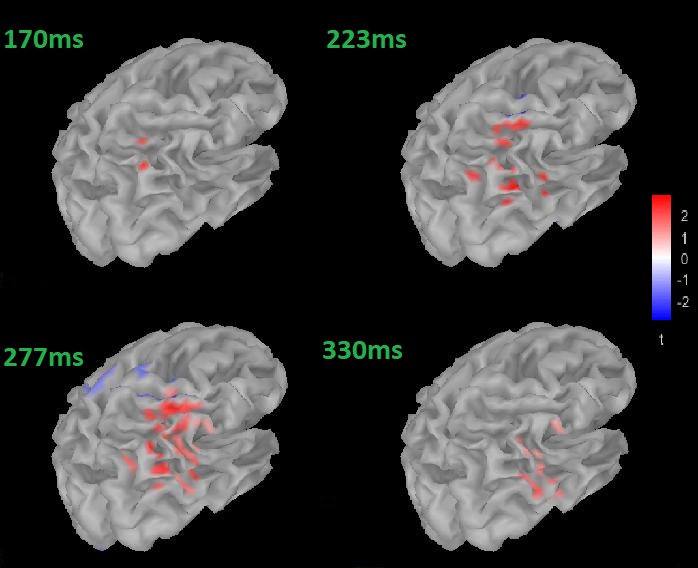}
     \caption{\normalsize{wMNE}}
  \end{subfigure}
  \begin{subfigure}[b]{0.45\textwidth}
    \includegraphics[width=\linewidth]{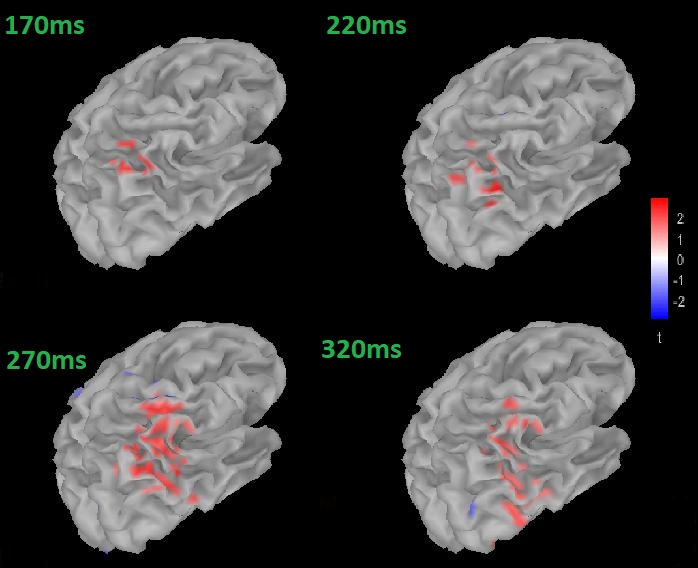}
    \caption{\normalsize{dSPM}}
  \end{subfigure}
  \begin{subfigure}[b]{0.45\textwidth}
    \includegraphics[width=\linewidth]{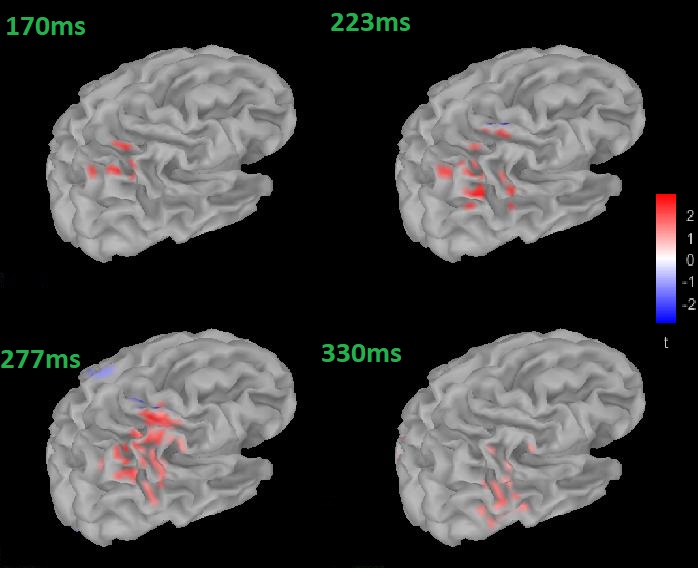}
    \caption{\normalsize{sLORETA}}
  \end{subfigure}
  \caption{ET vs CT: IPL of the right hemisphere activates more for the ET group at the time interval from 160 to 360 ms.}
  \label{figure2}
\end{figure}

Having guaranteed that the T-test can be applied, we proceeded with the comparisons ET vs CT and ENT vs CNT. The null hypothesis $H_{0}$ for both comparisons was that the two groups have equal means regarding the emerged spatio-temporal activation at the source level. Again, we performed ~$3.E^6$ simultaneous two-sample T-tests with $p<0.05$ level of significance. FDR was used to deal with the multicomparison problem. Another constraint that we added to avoid spurius and random effects was the one of the minimum duration of the activations. Thus, we excluded all the signals that were statistically significantly for time intervals less than 50 ms. This statistical analysis revealed that all methods gave relatively similar results. The pair T-test between ENT and CNT revealed a statistically significant difference in the time-range of 170ms-230ms. In this time-range, the Superior Parietal Lobe (SPL) was activated more in the ENT group; the activation of the SPL was mostly at the right hemisphere (figure \ref{figure1}). The pair T-test between ET and CT revealed a statistically significant difference in the time-range of 160ms-360ms. In this time-range the Inferior Parietal Lobule (IPL) was activated more in the ET group (only at the right hemisphere) (figure \ref{figure2}).

The above findings are summarized in Table \ref{results}.

\begin{table}[h!]
\centering
\resizebox{\columnwidth}{!}{%
\begin{tabular}{ccccccc}
\textbf{Comparison} & \textbf{Time period} & \textbf{\begin{tabular}[c]{@{}c@{}}Group activated\\ more\end{tabular}} & \textbf{Brain region} & \textbf{Brodman area} & \textbf{Number of voxels} & \textbf{Comments} \\ \hline
\multicolumn{1}{c|}{\textbf{ENT vs CNT}} & 170-230 & ENT & SPL & 7 & $\sim$80 & \begin{tabular}[c]{@{}c@{}}Only at the right\\ hemisphere\end{tabular} \\ \hline
\multicolumn{1}{c|}{\textbf{ET vs CT}} & 160-360 & ET & IPL & 22, 39-40 & $\sim$170 & \begin{tabular}[c]{@{}c@{}}Only at the right\\ hemisphere\end{tabular}
    \end{tabular}
    }%
\caption{Analytical presentation of the differences between the comparisons ET vs CT and ENT vs CNT.}
\label{results}
\end{table}

\newpage

\section{Conclusions}

Sensor-level analysis does not provide information about the actual sources that are involved in brain activity. From a mathematical point of view, the problem of source localization from scalp recordings is an inverse ill-defined problem and as such, different types of regularization approaches may in principle result to different solutions. Thus, especially for clinical assessment of brain neurological disorders such as epilepsy, there is a need for comparing and assessing the robustness of such methods. \\
This is the first study to perform a comparative analysis of three numerical methods, namely the sLORETA, wMNE and dSPM to identify differences at the source level between healthy children and children with well-controlled epilepsy (i.e. in the absence of seizures) during a working memory task . In the absence of anatomical MRI scans, we used the publicy ``Neurodevelopmental MRI database" that provides age-specific average MRI templates. Our analysis shows that all three methods yield essentially the same results, thus providing adequate confidence for our findings. More specifically, our analysis revealed consistent differences between the two groups in the parietal lobes. Importantly, our findings are in line with other studies investigating abnormalities of the brain function due to epilepsy with the use of anatomical MRI,  fMRI and EEG recordings \cite{Berryhil2008,Jones2012,Osaka2007, Li2017, Tseng2018, Herman2006, Hu2017, Kipp2012, Michels2010,Besenyei_2012}.  More specifically, regarding children with epilepsy, Besenyei et al. \cite{Besenyei_2012} used anatomical MRI and resting state EEG recordings to identify the abnormal brain activity in children with benign rolandic epilepsy. Using LORETA, they found an increase activity, compared to controls, in the temporal and inferior parietal lobule. Other studies that have investigated the abnormal activity in children with epilepsy in the presence of seizures have also pinpointed the importance of these areas. Clemens et al. (\cite{Clemens_2016} find increased activity in the superior perietal lobe  children with benign
childhood epilepsy with rolandic spikes, using  MRI scans and resting state EEG recordings. For the source localization analysis they used LORETA.\par
Importantly, our study and findings reveal also the importance and potential that originates from the use of publicly available scientific resources such as the ``Neurodevelopmental MRI" database, which allow to the researchers to re-analyse available neuroimaging data and investigate questions beyond the scope of the original studies. This carries, in principle, the potential to gain new insights without the need to perform new from scratch, time-consuming and expensive experiments.

\section*{Acknowledgments}
E.G. was supported by a Ph.D. fellowship by the Department of Mathematics and Applications, University of Naples Federico II and I.G. was supported by a Ph.D. fellowship by the National Technical University of Athens.

\section*{Conflict of interest}
All authors declare no conflicts of interest in this paper



\end{document}